\documentclass[preprint,amsmath,amssymb]{revtex4}

\usepackage{dcolumn}
\usepackage{bm}
\usepackage{longtable, color}
\usepackage{multirow}
\usepackage{subfigure}
\usepackage{pslatex}
\usepackage{courier}
\usepackage[dvips]{epsfig}
\usepackage[dvips]{graphicx}

\newcommand{\Eq}[1]{Eq. (\ref{#1})}
\newcommand{\Eqs}[1]{Eqs.\,(\ref{#1})}
\newcommand{\Fig}[1]{Fig.\,\ref{#1}}

\def\Sch{Schr\"{o}dinger\ }

\def\la{\langle}
\def\ra{\rangle}

\def\a0{$a_{0}$}

\begin{document}

\title{Ultralong Rydberg Cs$_2$ Molecules Investigated by Combined {\it ab initio} Calculations and Perturbation Theory}

\author{Xiaomeng Liu}
\affiliation{State Key Laboratory of Quantum Optics and Quantum
Optics Devices, Laser Spectroscopy Laboratory, Shanxi University,
Taiyuan 030006, China}
\author{Yonggang Yang}
\email{ygyang@sxu.edu.cn} \affiliation{State Key Laboratory of
Quantum Optics and Quantum Optics Devices, Laser Spectroscopy
Laboratory, Shanxi University, Taiyuan 030006, China}
\author{Jianming Zhao}
\affiliation{State Key Laboratory of Quantum Optics and Quantum
Optics Devices, Laser Spectroscopy Laboratory, Shanxi University,
Taiyuan 030006, China}
\author{Liantuan Xiao}
\affiliation{State Key Laboratory of Quantum Optics and Quantum
Optics Devices, Laser Spectroscopy Laboratory, Shanxi University,
Taiyuan 030006, China}
\author{Suotang Jia}
\affiliation{State Key Laboratory of Quantum Optics and Quantum
Optics Devices, Laser Spectroscopy Laboratory, Shanxi University,
Taiyuan 030006, China}

\date{\today}

\begin{abstract}
Vibrational properties of ultralong Rydberg Cs$_2$ molecules are
investigated on corresponding potential curves obtained by
perturbation theory. The Rydberg Cs$_2$ molecules are associated by
a Rydberg Cs($nS/nP$) atom $(n=30-70)$ and a ground state Cs($6s$)
atom. The starting point for the perturbation treatment of
corresponding Rydberg molecular potential curves is to generate
accurate atomic Rydberg states from realistic {\it ab initio}
effective core potential. The calculated results have similar
characteristics with available experimental and theoretical
investigations on Rydberg Rb$_2$ molecules. And this is the first
time that Rydberg molecules are studied at the {\it ab initio}
level.
\end{abstract}

\maketitle

\section{Introduction}\label{intro}
Owing to the rapid development on ultracold atoms and
molecules\cite{95Science269198,03Science299232,Carr09New11}, a novel
type of molecules, Rydberg molecules become accessible to nowadays
experiments\cite{09Nature4581005,09NP581}. A Rydberg molecule is
typically formed by the unusually long-range interaction between a
Rydberg atom and a ground state (or another Rydberg) atom.
Consequently Rydberg molecules are classified into two types
according to their formation. In the first case, a macrodimer
Rydberg
molecule\cite{03PRl183002,02PRL88133004,05JPB38S295,06PRA74020701}
associated by two Rydberg atoms (A*+B*) has a size larger than 1
$\mu$m and in the second case a butterfly or trilobite like Rydberg
molecule\cite{00PRL852458,02PRA66042709,02jpb53l199,09nature975greene}
associated by a Rydberg atom and a ground state one (A*+B) has a
typical size of about $10^3$ $a_0$ (Bohr radius). Since the first
calculation of Rydberg molecular potential curves in 2000 by Greene
and coworkers\cite{00PRL852458}, Rydberg molecules have been
extensively investigated both theoretically and
experimentally\cite{09PRA80052506,10PRL105163201,11JPB184004,11PRA83050501}
due to their extraordinary properties such as huge size, large
dipole moment, weak bounding energy and long lifetime. In
particular, in some fast developing area of quantum physics, Rydberg
molecules are more suitable than Rydberg atoms for quantum
 manipulation with electric
 field\cite{11Science3341110,12PRA85052511}.

The A*+B* and A*+B types of Rydberg molecules have been postulated
in Refs. \cite{02PRL88133004} and \cite{00PRL852458} respectively.
Soon experimental
evidence\cite{03PRL90143002,03PRL91183002,04PRL93163001} has been
found for the A*+B* type of Rydberg molecules. Later on experimental
verification of the A*+B type of Rydberg
molecules\cite{06PRL97233002,09PRA80052506} has been reported. In
2009 both types of Rydberg molecules have been reported to be
directly observed experimentally\cite{09Nature4581005,09NP581}. In
Ref. \cite{09Nature4581005} Bendkowsky and coworkers have reported
the measured spectra of Rydberg molecular states formed by Rb($5s$)
and Rb($ns$) with high principle quantum number $n=34,\cdots,40$.
While in Ref. \cite{09NP581} electric field induced Cs Rydberg atom
macrodimers bound at internuclear separations of $3-9$ $\mu$m are
investigated, focusing on $(63d)+(65d)$, $(64d)+(66d)$,
$(65d)+(67d)$ and $(66d)+(68d)$ pairs. Soon after that Rydberg
trimers\cite{10PRL105163201} have also been created by a single-step
photoassociation. For a better understanding of the mechanism and
further manipulations of Rydberg molecules, coherent transfer of
initially free pairs of Rb($5s$) atoms to Rydberg molecules (Rb*+Rb)
have been demonstrated experimentally\cite{10np6970}. Very recently
permanent electric dipole moment of a homonuclear Rydberg molecule
has been observed\cite{11Science3341110} to has the order of 1 Debye
even for highly symmetric case of Rb($ns$)+Rb($5s$) which further
pave the way for possible experimental manipulation of this type of
molecules.

From theoretical point of view, there have been some calculations of
the interaction potential curves and vibrational properties of
Rydberg molecules
\cite{00PRL852458,02PRA66042709,02PRL88133004,05JPB38S295,06PRA74020701,11PRA83050501,12PRA85052511}
 most of which focusing on A*+B* type.
The available theoretical calculations have played significant roles
in the experimental realization and interpretation of molecular
Rydberg states. In Ref. \cite{11PRA83050501} the photoassociation
rates of different bound states have been investigated for Rb
Rydberg microdimers which implies potential application to quantum
information processing. Very recently the electric field control of
triatomic Rydberg molecules have been reported\cite{12PRA85052511}
which shows their sensitivity to external electric field and
non-adiabatic transitions between neighboring Rydberg molecular
potential curves. For A*+B type of Rydberg molecules, in general the
available theoretical investigations\cite{00PRL852458,02PRA66042709}
have adopted $s$ wave scattering or perturbation treatment based on
Rydberg states from certain empirical electron-core potential for
the Rydberg atom. Therefore accurate description of the Rydberg atom
is essential for the accuracy of the Rydberg molecular potential
curve no matter by scattering or perturbation treatments, in
particular when pursuing quantitative agreement with experiments. To
meet this challenge, the present work will adopt combined {\it ab
initio} calculations and a long range analytical form for better
description of the electron-core potential of a Rydberg atom to
obtain accurate potential energy curves and vibrational properties
of a Rydberg molecule. According to our best knowledge this is the
first time that Rydberg molecules are studied at the {\it ab initio}
level.

In this work ultralong Rydberg Cs$_2$ molecular potential curves
dissociating into Cs($nS/nP$)+Cs($6s$) and corresponding vibrational
levels will be investigated $(n=30-70)$. Details of the methods
including combined {\it ab initio} and analytical description of Cs
Rydberg atoms, perturbation treatment of Cs$_2$ molecular potential
curves and diagonalization of Cs$_2$ vibrational Hamiltonian in
Fourier grid discrete variable representation\cite{89JCP9115} are
given in Sections \ref{Hatom}, \ref{Hmole} and \ref{Diag},
respectively. Then the Rydberg Cs$_2$ molecular potential curves,
vibrational levels, equilibrium internuclear distances and ground
state binding energies are discussed in Section \ref{resultdiss}.
Finally the conclusions of the present work are summarized in
Section \ref{sec:conc}.

\section{Methods}\label{method}
\subsection{Hamiltonian and Wavefunction of Cs Atom}\label{Hatom}
The probability density of Rydberg electron is crucial for the
potentials of the long-range Rydberg molecules. Therefore the
Rydberg states of a cesium atom need to be accurately solved as a
prerequisite for studying Rydberg Cs$_2$ molecules. Based on
central-field approximation, the cesium atom can be described by a
single electron model with the core-electron interaction using an
effective core potential. The Hamiltonian of a cesium atom reads
\begin{equation}\label{ww-01}
    H_{\rm atom}=-\frac{\hbar^{2}}{2m}\nabla^{2}+V_{\rm atom}(r),
\end{equation}
where $V_{\rm atom}(r)$ is the effective core potential. For large
$r$ the effective core potential has standard long range asymptotic behavior
as
\begin{equation}\label{vatom}
    V_{\rm atom}(r)=- \frac{e^2}{4\pi\varepsilon_0 r}-\frac{e^2}{(4\pi\varepsilon_0)^2}\cdot \frac{\alpha_c}{2r^4}
\end{equation}
with $\alpha_c$ being the Cs$^+$ core polarizability. While for
small $r$ the potential becomes considerably more complicated
consequently the short range potentials are directly obtained by
{\it ab initio} calculations using Gaussian09 suit of
programs.\cite{gaussian09} The CCSD(T) method, coupled cluster
including single and double excitations and triple excitations by
perturbation, is adopted to get accurate energies. For modern
quantum chemistry calculations the CCSD(T) method has been widely
used for accurate energy calculations of small
molecules\cite{pes2005h5o2,yang12jpca} due to its balance between
speed and accuracy. In this work all {\it ab initio} energy
calculations are performed with Gaussian09 at CCSD(T)/QZVP level of
theory and full correlation between all the electrons explicitly
involved in the QZVP basis set has been calculated.

Having the Cs atom Hamiltonian at hand, the single electron \Sch
equation can be solved to get the valence electron wavefunctions.
The total wavefunction $\Psi_{\rm atom}(\mathbf{r})$ can be
separated into the radial and angular parts as $\Psi_{\rm
atom}(\mathbf{r})=\Psi_{\rm
atom}(r,\theta,\varphi)=\mathfrak{R}_{nl}(r)Y_{l}^{m}(\theta,\varphi)$
with $Y_{l}^{m}(\theta,\varphi)$ being the well known spherical
harmonics. While the radial wavefunction $\mathfrak{R}_{nl}(r)$ can
be solved by the numerical methods such as Numerov
method\cite{ThijssenCP}. Taking $X(r)=r\mathfrak{R}_{nl}(r)$, the
radial \Sch equation can be written as standard Numerov form
    $\frac{d^2X}{dr^2}=-f(r)X$,
where $f(r)=\frac{2m}{\hbar^2}\left[E_{nl}-V_{\rm
atom}(r)\right]-\frac{l(l+1)}{r^2}$ with the corresponding Cs atom
eigenenergy $E_{nl}$. From the quantum defect theory, the fine
structure eigenenergy of Cs atom $E_{nlj}$ can be obtained with the
quantum defects reported in Ref. \cite{82PRA262733}. With $E_{nlj}$
it is straightforward to derive the eigenenergy $E_{nl}$.
Discretizing $r$ by $r_{j}=(N-j+1)\cdot h$, where $h$ is the step
size and $N$ is the number of total grids, the Numerov integration
is written as
\begin{equation}\label{num-1}
    X_{i+1}=\frac{\left(2-\frac{5h^{2}}{6}f_{i}\right)X_{i}-\left(1+\frac{h^2}{12}f_{i-1}\right)
    X_{i-1}}{1+\frac{h^{2}}{12}f_{i+1}}+O(h^6).
\end{equation}
The final integration is being done starting from the long range limit where the eigenfunction has
asymptotic form  $X_{1,2}=\exp\left(-\sqrt{-2mE_{nl}}r_{1,2}/\hbar\right)$.

\subsection{Perturbation Treatment of Rydberg Cs$_2$ Molecular Potential Energy Curves}\label{Hmole}
\begin{figure}[htbp]
\centering
\includegraphics [angle=0,scale=0.75] {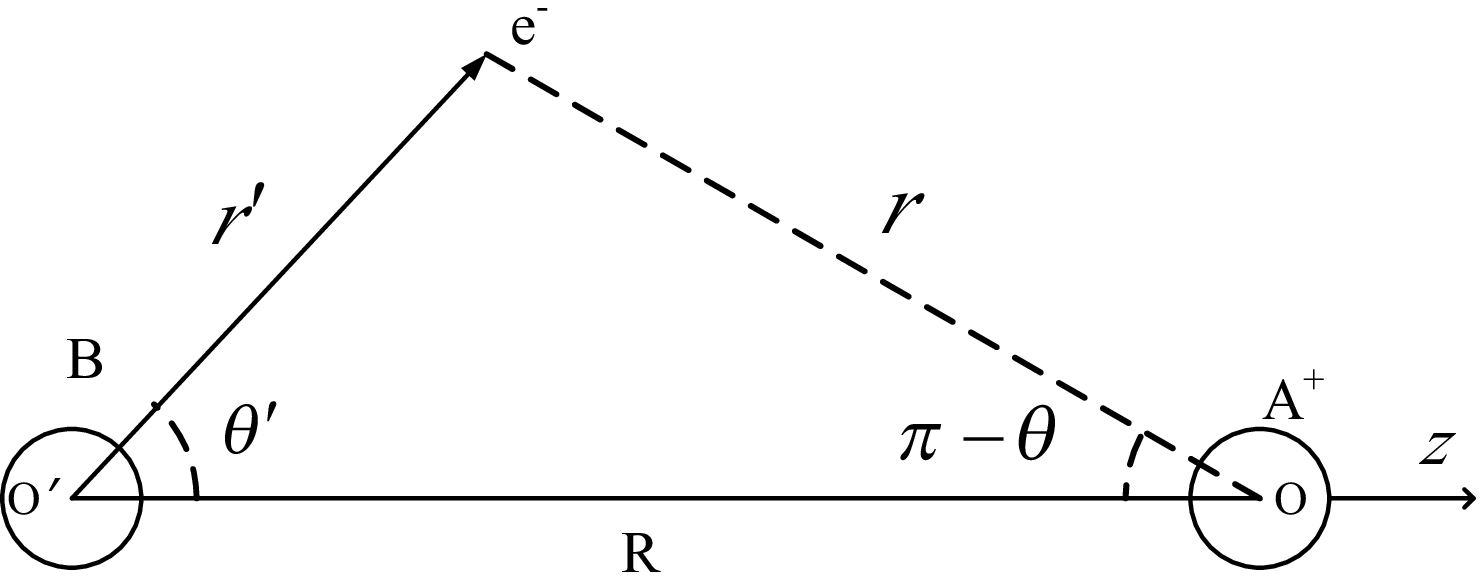}
\caption{Liu et al} \label{fig:mole}
\end{figure}

A Rydberg Cs$_2$ molecule, formed by a ground state Cs($6s$) atom and a Rydberg Cs($nl$) atom,
can be described by a three-body model shown in Fig. \ref{fig:mole}.
The $A^+$ and $e^-$ are the core and the valence electron of the Rydberg Cs($nl$) atom respectively
and $B$ is the ground state Cs($6s$) atom.
To be consistent with the above mentioned atomic part, $r$ is the distance between the valence electron
and the Cs$^+$ core, while $r'$ is the distance between the valence electron and the Cs($6s$) atom.
And the two Cs atoms are separated by a large distance $R$.
The total Hamiltonian can be obtained by taking into account of all the two-body
interactions, namely, the molecular
Hamiltonian of long-range Rydberg Cs$_2$ molecule can be written as
\begin{equation}\label{ww-02}
    H_{\rm mol}=H_{\rm atom}+V_{AB}(R)+V_{Be}(r')+T_{AB},
\end{equation}
where $V_{Be}$ ($V_{AB}$) is the interaction potential between $B$
and $e^-$ ($A^+$ and $B$)
\begin{subequations}\label{ww-03}
\begin{equation}
  V_{Be}(r')=-\frac{e^{2}}{(4\pi\varepsilon_{0})^{2}}\frac{\alpha_{a}}{2r'^{4}}
  \end{equation}
  \begin{equation}
  V_{AB}(R)=-\frac{e^{2}}{(4\pi\varepsilon_{0})^{2}}\frac{\alpha_{a}}{2R^{4}}
\end{equation}
\end{subequations}
and $T_{AB}$ is the kinetic energy of the two Cs atoms. Here
$\alpha_{a}$ is the polarizability of a ground state Cs($6s$) atom
and \Eqs{ww-03} is valid for long range interactions. Similar to the
atomic effective core potential $V_{\rm atom}(r)$, the short range
behavior of $V_{Be}(r')$ is complicated consequently it is directly
obtained by {\it ab initio} calculations. While the short range
behavior of $V_{AB}(R)$ is not needed in the present work since the
prime interest of the present work is the long range Rydberg Cs$_2$
molecules.

Using Born-Oppenheimer approximation, the molecular \Sch equation
can be separated into two parts and the molecular potential energy
curve can be obtained from the electronic \Sch equation
\begin{subequations}
\label{ww1}
\begin{equation}
  \left(H_{\rm atom}+V_{Be}(r')\right)
\psi_{el}({\bf r,R})=E_{el}(R)\psi_{el}({\bf r,R}) \label{ww1-1}
\end{equation}
\begin{equation}
V_{\rm mol}(R)=E_{el}(R)+V_{AB}(R) \label{ww1-2}
\end{equation}
\end{subequations}
In this work, the electronic \Sch equation \Eq{ww1-1} is solved by
perturbation theory. For ultralong Rydberg molecules the interaction
potential $V_{Be}(r')$ has small matrix elements between Rydberg
states of $H_{\rm atom}$ due to short range characteristics of
$V_{Be}(r')$ and widely delocalized Rydberg electron density, which
can also be seen from the results section. Consequently it is valid
to set $H_{\rm atom}$ as zero order Hamiltonian and the interaction
potential $V_{Be}$ as perturbation. Since $V_{Be}$ only leads to
couplings between the originally non-degenerate atomic states, the
corresponding potential curve to the first order perturbation theory
reads
\begin{eqnarray}
V_{\rm mol}(R)=V_{AB}(R)+E_{el}(R)=V_{AB}(R)+\langle \Psi_{\rm atom}|V_{Be}(r')|\Psi_{\rm atom}\rangle \label{ww-wei3}
\end{eqnarray}
\Eq{ww-wei3} is the potential energy with respect to the
corresponding dissociation limit of the two Cs(6s)+Cs($nl$) atoms.
The final numerical integrations for the perturbation term are
calculated by
\begin{eqnarray}
    \langle\Psi_{\rm atom}|V_{Be}(r')|\Psi_{\rm atom}\rangle=\iiint V_{Be}(r')\mathfrak{R}_{nl}^{2}(r)Y_{lm}^{*}
    (\theta,\varphi)Y_{lm}(\theta,\varphi)r^{2}\sin\theta dr d\theta d\varphi \nonumber \\
=2\pi \iint V_{Be}(r')\mathfrak{R}_{nl}^{2}(r)Y_{l0}^{2}\left(\pi-\arccos\left(\frac{R-r'\cos\theta'}
    {r}\right),0\right)r'^{2}\sin\theta' dr'd\theta' \nonumber
\end{eqnarray}
for different $R$ with $r=\sqrt{R^{2}+r'^{2}-2Rr'\cos\theta'}$ to properly treat the singular point of
the interaction $V_{Be}(r')$ at $r'=0$ (which behave as $-1/r'$).

\subsection{Numerical Diagonalization of Cs$_2$ Molecular Vibrational Hamiltonian} \label{Diag}
The vibrational states of Rydberg Cs$_2$ molecules can be obtained by diagonalization of the
molecular radial/vibrational Hamiltonian.
Having the knowledge of the molecular potential energy $V_{\rm mol}(R)$ the molecular rovibrational
Hamiltonian reads
\begin{equation}\label{rovibH}
H_{\rm mol}^{rovib}=T_{AB}+V_{\rm mol}(R)=-\frac{\hbar
^2}{2\mu}\frac{1}{R^2}\frac{\partial}{\partial
R}R^2\frac{\partial}{\partial R}+V_{\rm
mol}(R)+\frac{\hat{J}^2}{2\mu R^2}
\end{equation}
with the reduced mass $\mu=\frac{1}{2}m_{Cs}$ and the total angular
momentum $\hat{J}^2$. The total rovibrational wavefunction can be
written as $\Psi_{\rm mol}(\mathbf{R})=\mathbb{R}_{\nu J}(R)Y_J^m$
with $\mathbb{R}_{\nu J}(R)$ being the radial/vibrational
wavefunction and $Y_J^m$ the spherical harmonics as mentioned in
previous section. The corresponding rovibrational energy level is
$E_{\nu J}$ with vibrational quantum number $\nu$ and rotational
quantum number $J$. For a given angular momentum $J$, \Eq{rovibH}
can be diagonalized to give eigenenergy $E_{\nu J}$ and
radial/vibrational wavefunction $\mathbb{R}_{\nu J}(R)$. Numerical
diagonalization of \Eq{rovibH} is performed in a Fourier grid
discrete variable representation\cite{89JCP9115,MM10ma} to give
vibrational energy $E_{\nu 0}$ and corresponding vibrational
wavefunction.

\section{Results and Discussion}\label{resultdiss}
\subsection{The Atomic Wavefunctions of Cs$(ns/np)$ Rydberg States}
\begin{figure}[htbp]
\centering
\includegraphics [angle=-90,scale=0.5] {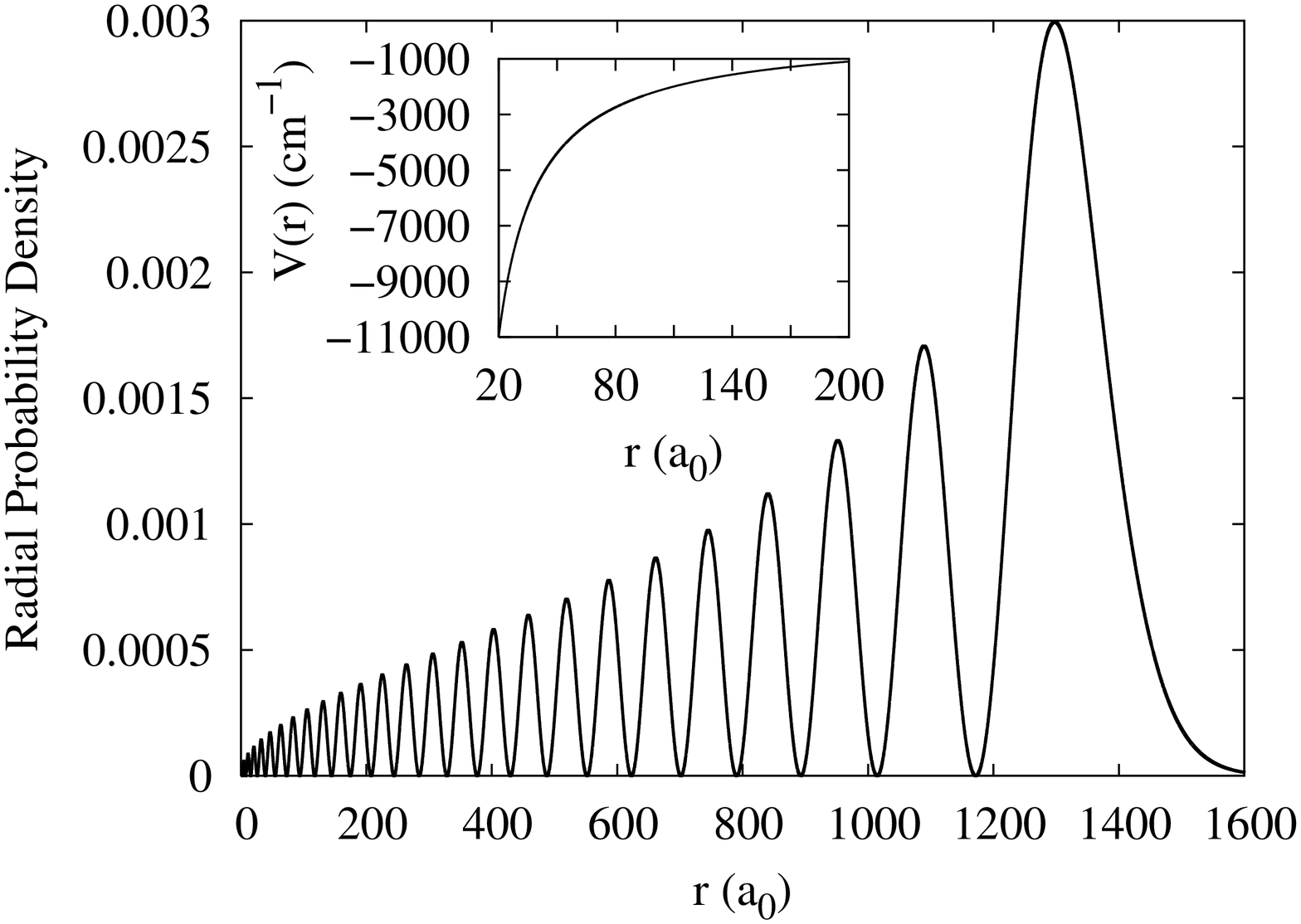}
\caption{Liu et al} \label{fig:r30-1}
\end{figure}

A typical Rydberg electron density is shown in \Fig{fig:r30-1} (for
Cs($30p$) state as an example). In the inset of \Fig{fig:r30-1} the
Cs effective core potential $V_{\rm atom}(r)$ is also shown. The
effective core potential $V_{\rm atom}(r)$ is calculated by
Gaussian09 suit of programs for $r<50\ a_0$ with a step size of
$0.1\ a_0$ at CCSD(T)/QZVP level of theory. While for $r>50\ a_0$
the $V_{\rm atom}(r)$ is first calculated by Gaussian09 at the same
level of theory then fitted to \Eq{vatom}. The fitted core
polarizability for Cs$^+$ is $\alpha_c=-16.0\ a_{0}^{3}$  with a
relative fitting error of less than 0.01\%. The calculated core
polarizability agree well with the experimental value of -15.5443
$a_{0}^{3}$ reported in Ref. \cite{80PRA222672}. For further
calculations the experimental core polarizability
$\alpha_c=-15.5443\ a_{0}^{3}$ is aopted for $r>50\ a_0$. The {\it
ab initio} calculated effective core potential for $r<50\ a_0$ and
the analytical one for $r>50\ a_0$ are connected smoothly as shown
in  the inset of \Fig{fig:r30-1}, which generates the full effective
core potential $V_{\rm atom}(r)$ used for the calculations of atomic
wavefunctions.

The $ns$ and $np$ states of Cs atom wavefunctions are calculated by
numerical integration of \Eq{num-1} with a step size of $0.1\ a_0$
and initial condition of the wavefunction at $r>2n^2\ a_0$ which
obey asymptotic form
$\mathfrak{R}_{nl}(r)\propto\frac{1}{r}\exp\left(-\sqrt{-2mE_{nl}}r/\hbar\right)$.
As can be seen from \Fig{fig:r30-1} the Rydberg electron density is
highly oscillating. The calculated most probable radius for the
valence electron of the Cs($30p$) state is 1297.7 $a_0$ which is the
typical order of magnitude of the size of a Rydberg atom. For the
Cs($30s$), Cs($60s$) and Cs($60p$) states, the calculated most
probable radii are 1250.9 $a_0$, 5989.4 $a_0$ and 6093.0 $a_0$,
respectively. The size of a Rydberg atom increases rapidly with
increasing principle quantum number $n$ and slightly increases with
increasing angular momentum quantum number $l$, both due to
increasing eigenenergy $E_{nl}$.

\subsection{The Potential Energy Curves of Cs$(ns/np)$+Cs$(6s)$ Rydberg Molecules}
Having the wavefunctions of Cs atomic Rydberg states at hand, the
corresponding Rydberg Cs$_2$ molecular potentials $V_{\rm mol}(R)$
can be obtained according to \Eq{ww-wei3} with the knowledge of the
interaction potentials $V_{Be}(r')$ and $V_{AB}(R)$. The long range
behaviors of $V_{Be}(r')$ and $V_{AB}(R)$ can be well described by
the ground state polarizability $ \alpha_{a}$ of Cs atom according
to \Eq{ww-03}. The Cs atom polarizability $ \alpha_{a}$ has been
obtained in the same way as the above mentioned core polarizability
$ \alpha_{c}$, namely, by fitting of the {\it ab initio} interaction
energy $V_{Be}(r')$ of $r'>50\ a_0$ calculated by Gaussian09 suit of
programs at CCSD(T)/QZVP level of theory. The fitted ground state
cesium atom polarizability is $ \alpha_{a}=-416.3\ a^3_{0}$ with a
relative fitting error of less than 0.002\%. The atom polarizability
$ \alpha_{a}$ is significantly larger than the core polarizability $
\alpha_{c}$ since the outmost $6s$ valence electron of the Cs atom
can be easily polarized. For $r'<50\ a_0$, $V_{Be}(r')$ has been
calculated by Gaussian09 suit of programs at CCSD(T)/QZVP level of
theory with a step size of $0.1\ a_0$.

\begin{figure}[htbp]
\centering
\includegraphics [angle=-90,scale=0.5] {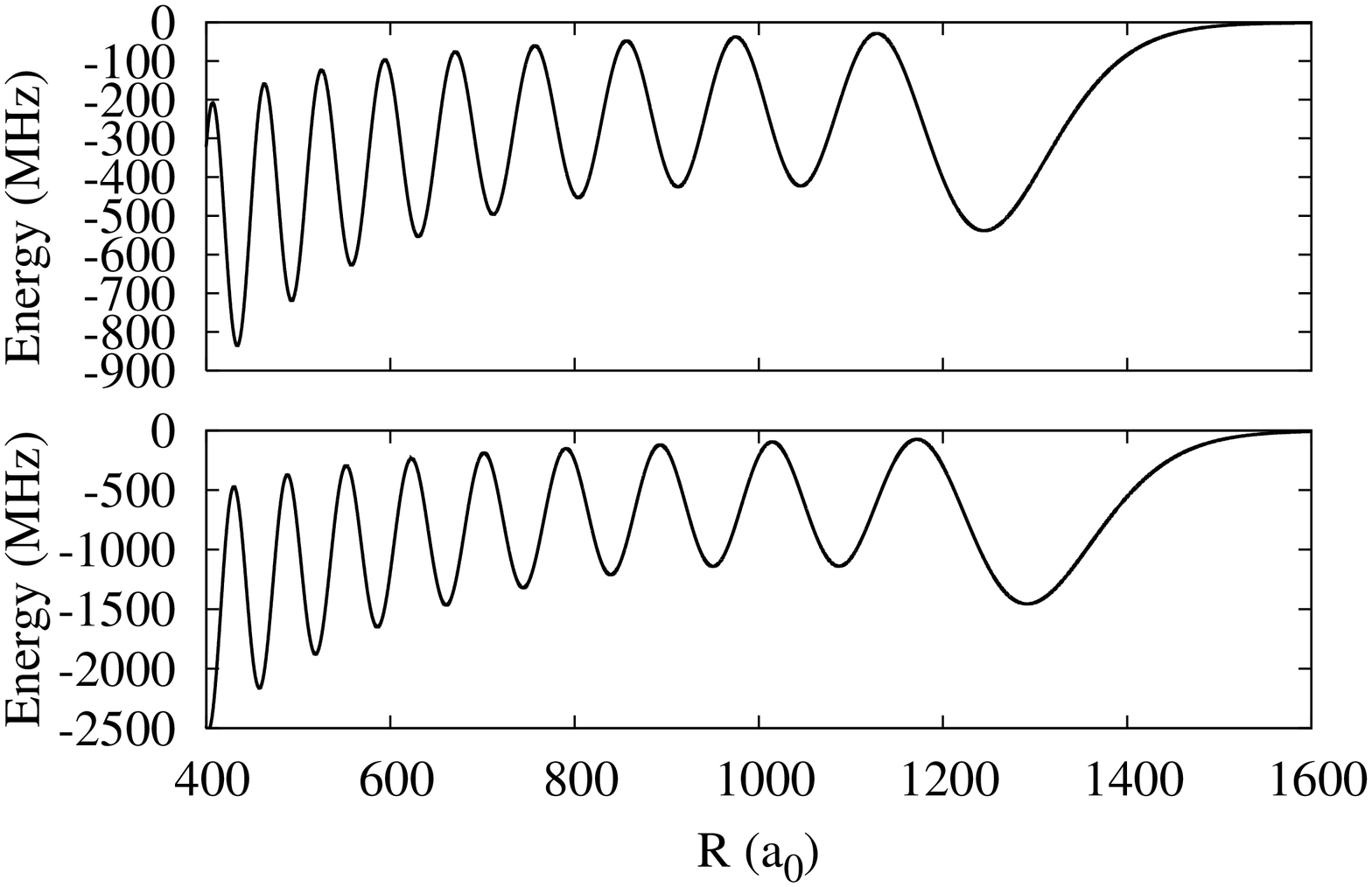}
\caption{Liu et al} \label{fig:e30-1}
\end{figure}

The calculated potential energy curves of Rydberg Cs$_2$ molecules
are shown in \Fig{fig:e30-1} for Cs($30s/30p$)+Cs($6s$) and
\Fig{fig:e60-0} for Cs($60s/60p$)+Cs($6s$). The potential curves
also oscillate due to the oscillating characteristics of the atomic
Rydberg electron density. For details, the potential curves
dissociating into $6s$ ground state and $ns$ ($np$) Rydberg state
are shown in the upper (bottom) panels of \Fig{fig:e30-1} and
\Fig{fig:e60-0}. Due to increasing size of a Rydberg atom with
angular momentum quantum number $l$, each bottom panel ($np$+$6s$)
has a larger equilibrium internuclear distance for the outmost
potential well than that of the upper panel ($ns$+$6s$). In general
the Rydberg molecules can be associated to the outmost potential
well more efficiently than other potential wells. As far as the
outmost potential well is concerned, the dissociation energy $D_e$
of the bottom panel ($np$+$6s$) is found to be considerably larger
than the upper panel ($ns$+$6s$) which reflects stronger interaction
between $np$ and $6s$ states than the one between $ns$ and $6s$
states for a given principle quantum number $n$. The calculated
results for Cs*+Cs agree with the reported trends for Rb*+Rb in Ref.
\cite{00PRL852458}.

Details of the dissociation energies $D_e$ and equilibrium
internuclear distance $R_{min}$ for the outmost potential wells of
Rydberg Cs$_2$ molecules associated from different atomic states are
shown in Table \ref{tab-nw}. Following the same symbols adopted
throughout the present work, a Rydberg Cs$_2$ molecule is formed
starting from a ground state Cs ($6s$) atom and a Rydberg state Cs
($nl$) atom. Detailed investigations are performed for the principle
quantum number $n$ varying from 30 to 70. As can be seen, the
dissociation energy $D_e$ decreases almost in an exponential way
with increasing principle quantum number $n$ of the Rydberg atom
which agrees with the experimentally observed trends for Rb*+Rb in
Ref. \cite{09Nature4581005}. This is because the Rydberg electron
density becomes more delocalized as the principle quantum number $n$
becomes larger. As a result the outmost maximum of the electron
density, namely the most probable radial probability, decreases with
increasing $n$ which weakens the interaction between the two Cs
atoms. While the molecular size, which is reflected by $R_{min}$ for
the outmost potential well, increases according to the well known
$n^2$ rule. By fitting the data reported in Table \ref{tab-nw} one
finds the relations $R_{min}=1.65n^2$ for $ns$+$6s$ Rydberg Cs$_2$
molecules and $R_{min}=1.68n^2$ for $np$+$6s$ ones, which are quite
close to each other.

\subsection{Vibrational States of Cs($ns/np$)+Cs($6s$) Rydberg molecules}
To further understand properties of the ultralong Rydberg Cs$_2$
molecules, the molecular vibrational energy levels and wave
functions are calculated based on the above mentioned potential
energy curves. As mentioned in the method section, numerical
diagonalization of \Eq{rovibH} is performed in a Fourier grid
discrete variable representation\cite{89JCP9115} to give vibrational
energy $E_{\nu 0}$ and corresponding vibrational wavefunction. Table
\ref{tab-1} shows some vibrational energies $E_{\nu 0}$ of the
Rydberg Cs$_2$ molecule dissociating into Cs($30s$) and Cs($6s$)
states. A total of 15925 grids equally spaced from R=107.5 $a_0$ to
R=1699.9 $a_0$ with a step of 0.1 $a_0$, which covers 17 potential
wells, are adopted for numerical diagonalization. The
diagonalization finds 58 vibrational bound states among which 10
states are predominately located in the outmost potential well. Also
shown in Table \ref{tab-1} are the average internuclear distance
$\la R\ra$ and variance $\la\Delta R\ra=\sqrt{\la(R-\la R\ra)^2\ra}$
of each corresponding vibrational state. In general the Rydberg
molecules can be associated to the outmost potential well more
efficiently than other potential wells. Therefore Table \ref{tab-1}
shows all the vibrational states which are predominately located in
the outmost potential well. In this work all the vibrational
energies are reported with respect to the corresponding dissociation
energy. As can be seen the most deeply bound state in the outmost
potential well is located at -498.5 MHz with an average internuclear
distance $\la{R}\ra$ of 1246.6 $a_0$. Compared to the corresponding
$D_e=-538.2$ MHz a relatively large zero point energy of about 40
MHz is found and the equilibrium vibrational frequency in this
potential well is about 78 MHz according to the corresponding
fundamental excitation energy.

Similar calculations have been done for the Rydberg Cs$_2$ molecules
dissociating into Cs($30p$) and Cs($6s$) states. A total of 15251
grids equally spaced from R=174.8 $a_0$ to R=1699.8 $a_0$ with a
step of 0.1 $a_0$, which covers 15 potential wells, are adopted for
numerical diagonalization. In total 80 vibrational bound states are
found among which 17 ones are predominately located in the outmost
potential well. Table \ref{tab-2} shows all the 17 vibrational
states in the outmost potential well. The most deeply bound
vibrational states are found to be located at -1391.0 MHz with
average internuclear distance $\la{R}\ra$ of 1292.5 $a_0$. Compared
to the corresponding $D_e=-1455.9$ MHz the zero point energy is
about 65 MHz and the equilibrium vibrational frequency in this
potential well is about 128 MHz according to the corresponding
fundamental excitation energy.

In general, as the vibrational energy level increases the mean
distance $\la{R}\ra$ and variance $\la{\Delta R}\ra$ increase due to
approaching of the dissociation limit as can be seen in Tables
\ref{tab-1} and \ref{tab-2}. However, for the highest few
vibrational states in Tables \ref{tab-1} and \ref{tab-2}, they
already have considerable population in the inner potential wells
leading to quite large variances $\la{\Delta R}\ra$ and decrease of
the mean distances $\la{R}\ra$. Comparison between the above
mentioned vibrational energies and the ones obtained from
diagonalization of a relatively smaller Hamiltonian matrix which
only covers the outmost potential well reveals that the vibrational
levels in the outmost potential well can also be accurately obtained
by the latter calculation (diagonalizing a smaller matrix). The root
means square deviations between this two calculations are found to
be less than 0.02 MHz for both the lowest 7 states in Table
\ref{tab-1} and the lowest 12 states in Table \ref{tab-2}. If only
the well bounded (lowest few) states are considered the deviations
between the above mentioned two calculations are almost zero. To
check the convergence of the above calculations, the vibrational
energies in Tables \ref{tab-1} and \ref{tab-2} are further compared
with results from diagonalization of a larger Hamiltonian matrix
which covers the long range limit up to $R=4000$ $a_0$. The root
mean square deviations between the two calculations are about 0.03
MHz for the vibrational energies in Table \ref{tab-1} and 0.1 MHz
for the ones in Table \ref{tab-2}. Again the deviations are almost
zero if only the well bounded (lowest few) states are considered.

Based on the above mentioned error analysis, further investigations
on vibrational states of Rydberg Cs$_2$ molecules dissociating into
Cs($6s$) and Cs($nl$) with high principle quantum number $n$ are
performed by covering only the outmost potential well. For the
Cs($60s/60p$)+Cs($6s$) molecules 13583/12587 grids equally spaced
from R=5641.7/5741.3 $a_0$ to R=6999.9 $a_0$ with a step of 0.1
$a_0$ are adopted. Table \ref{tab-3} shows 3 vibrational bound
states in the outmost well for Cs($60s$)+Cs($6s$) case and 5
vibrational bound states in the outmost well for Cs($60p$)+Cs($6s$)
case. The corresponding zero point energy and equilibrium
vibrational frequency in the outmost potential well of
Cs($60s/60p$)+Cs($6s$) are 1.5/2.6 MHz and 2.7/4.9 MHz,
respectively.

\section{Conclusions}\label{sec:conc}
Ultralong range Rydberg molecules have been extensively studied in
the past decade due to its unique properties such as huge size,
large dipole moment and long lifetime. In the present work ultralong
Rydberg Cs$_2$ molecules formed by Cs$(ns/np)$+Cs$(6s)$ for
$(n=30-70)$ have been investigated focusing on molecular potential
curves and vibrational properties. The Rydberg molecular potential
curves are calculated by perturbation theory based on accurate
Rydberg atom wavefunctions. The Rydberg atom wavefunctions are
generated from realistic {\it ab initio} effective core potential
rather than some empirical effective core potential, which serves as
a good starting point for further calculations of potential energy
curves. The calculated potential energy curves have oscillating
nature due to the oscillation of the Rydberg electron density. The
investigated Rydberg Cs$_2$ molecular size increases with $n$
according to $R_{min}=1.65n^2$ for Cs$(ns)$+Cs$(6s)$ case and
$R_{min}=1.68n^2$ for Cs$(np)$+Cs$(6s)$ case. The dissociation
energy of the outmost potential well decreases almost exponentially
with increasing principle quantum number $n$ but increases
dramatically with increasing angular momentum quantum number $l$.
The largest binding energy of the vibrational ground states studied
in this work is 1391.0 MHz for Cs$(30p)$+Cs$(6s)$ case and quickly
decreases to a few MHz for $n=70$ case. According to our best
knowledge this is the first time that vibrational properties of
Cs*+Cs have been investigated quantitatively at the {\it ab initio}
level. Consequently the reported results are expected to play
important roles for future experiments on observation and
manipulation of Cs*+Cs kind of molecules. For each vibrational
levels predominately located in the outmost potential well, apart
from the vibrational energies, the mean values and variances of the
internuclear distances are also reported which may provide further
clues for the experimentalists.


%
\begin{acknowledgments}
This work was supported in part by 973 Program of China under Grant
No. 2012CB921603, the National Natural Science Foundation of China
under Grant No. 11004125, International science \& technology
cooperation program of China (2011DFA12490), National Natural
Science Foundation of China under Grant Nos. 10934004, 60978018,
11274209 and NSFC Project for Excellent Research Team (61121064).
\end{acknowledgments}

\clearpage

\bibliographystyle{jcp}
\bibliography{cs}

\clearpage

\clearpage
\section*{Figure Legends}
\subsubsection*{Fig.~\ref{fig:mole}:}
Rydberg molecule structure and coordinate system. The $A^+$ and
$e^-$ are the core and the valence electron of the Rydberg Cs($nl$)
atom respectively and $B$ is the ground state Cs($6s$) atom. See the
text for details.

\subsubsection*{Fig.~\ref{fig:r30-1}:}
The radial probability density $\rho(r)=|r\mathfrak{R}_{nl}|^2$ of
Cs$(30p)$ atom. The effective core potential $V_{\rm atom}(r)$ of
the Cs atom is shown in the inset.

\subsubsection*{Fig.~\ref{fig:e30-1}:}
Molecular potential curves for Cs$(30s)$+Cs$(6s)$ (Upper) and
Cs$(30p)$+Cs$(6s)$ (Bottom).

\begin{figure}[htbp]
\centering
\includegraphics [angle=-90,scale=0.5] {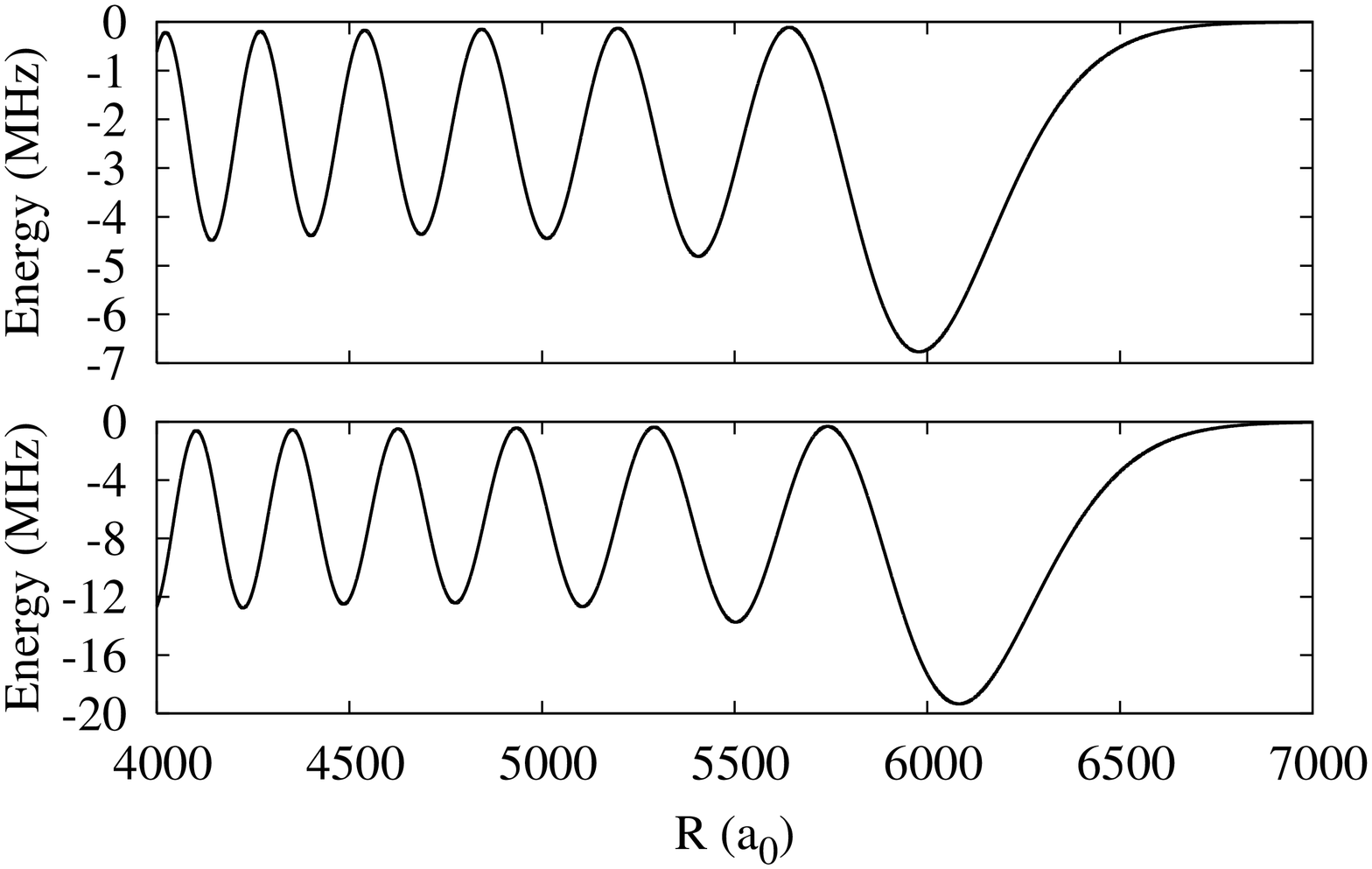}
\caption{Liu et al} \label{fig:e60-0}
\end{figure}

\subsubsection*{Fig.~\ref{fig:e60-0}:}
Molecular potential curves for Cs$(60s)$+Cs$(6s)$ (Upper) and
Cs$(60p)$+Cs$(6s)$ (Bottom).

\clearpage
\begin{table}
  \centering
  \caption{The calculated dissociation energy $D_e$ and equilibrium
internuclear distance $R_{min}$ for the outmost potential well of a
Rydberg Cs$_2$ molecules associated from a Rydberg Cs$(ns/np)$ atom
and a ground state Cs$(6s)$ atom.}\label{tab-nw}
  \begin{tabular}{c|c|c|c|c}
    \hline\hline
 & \multicolumn{2}{c|}{$ns+6s$} &\multicolumn{2}{c}{$np+6s$}\\ \hline
 $n$  &$R_{min}$ ($a_0$) & $D_e$ (MHz) &  $R_{min}$ ($a_0$) & $D_e$ (MHz)\\ \hline\hline
    30&1244.6& -538.2& 1291.4& -1455.9 \\ \hline
    35&1787.0& -197.3& 1843.2&  -542.7 \\ \hline
    40&2427.9&  -84.1& 2493.5&  -234.0 \\ \hline
    45&3167.4&  -40.0& 3242.5&  -112.5 \\ \hline
    50&4005.7&  -20.8& 4090.2&   -58.8 \\ \hline
    55&4942.7&  -11.5& 5036.8&   -32.8 \\ \hline
    60&5978.6&   -6.8& 6082.2&   -19.3 \\ \hline
    65&7113.5&   -4.2& 7226.5&   -11.9 \\ \hline
    70&8347.2&   -2.7& 8469.8&    -7.6 \\ \hline\hline
  \end{tabular}
\end{table}

\clearpage
\begin{table}
  \centering
  \caption{Vibrational properties of Cs$(30s)$+Cs$(6s)$ Rydberg molecules. $E_{\nu
0}$, $\la{R}\ra$ and $\la{\Delta R}\ra$ are the vibrational energy,
mean value and variance of the internuclear distance $R$,
respectively.}\label{tab-1}
  \begin{tabular}{c|c|c|c|c|c|c|c|c|c|c}
    \hline\hline
    $E_{\nu 0}$ (MHz) & -498.5 & -420.5 & -347.4 & -279.6 & -217.3 & -160.9 & -111.2 & -69.2 & -32.7 & -10.1 \\ \hline
    $\la{R}\ra$ $(a_0)$ & 1246.6 & 1250.8 & 1255.8 & 1261.5 & 1268.4 & 1276.6 & 1286.5 & 1285.9 & 1258.1 & 1328.0 \\ \hline
    $\la{\Delta R}\ra$ $(a_0)$ & 18.6 & 32.9 & 43.8 & 53.6 & 63.3 & 73.4 & 85.3 & 122.0 & 215.5 & 218.4 \\
    \hline\hline
  \end{tabular}
\end{table}
\clearpage
\begin{table}
  \centering
  \caption{Vibrational properties of Cs$(30p)$+Cs$(6s)$ Rydberg molecules. $E_{\nu
0}$, $\la{R}\ra$ and $\la{\Delta R}\ra$ are the vibrational energy,
mean value and variance of the internuclear distance $R$,
respectively.}\label{tab-2}
  \begin{tabular}{c|c|c|c|c|c|c|c|c|c}
    \hline\hline
    $E_{\nu 0}$ (MHz) & -1391.0 &-1262.7 & -1138.9 & -1019.8 & -905.5 & -796.2 & -692.1 & -593.4 & -500.4  \\ \hline
    $\la{R}\ra$ $(a_0)$ & 1292.5 & 1295.0 & 1297.7 & 1300.6 & 1303.8 & 1307.3 & 1311.1 & 1315.5 & 1320.4   \\ \hline
    $\la{\Delta R}\ra$ $(a_0)$ & 14.5 & 25.5 &33.4 & 40.2 & 46.5 & 52.4 & 58.3 & 64.1 & 70.1   \\ \hline\hline
    $E_{\nu 0}$ (MHz) &  -413.2 & -332.4 & -258.3 & -191.8 & -132.8 & -86.1 & -38.5 & -8.9 & \\ \hline
    $\la{R}\ra$ $(a_0)$ & 1325.9 &  1332.3 & 1339.2 & 1345.3 & 1351.0 & 1256.0 & 1268.8 & 1247.9 & \\ \hline
    $\la{\Delta R}\ra$ $(a_0)$ & 76.2 &  82.8 & 91.7 & 101.3 & 120.4 & 225.3 & 299.4 & 377.1 & \\ \hline\hline
  \end{tabular}
\end{table}
\clearpage
\begin{table}
  \centering
  \caption{Vibrational properties of Cs$(60s)$+Cs$(6s)$ and Cs$(60p)$+Cs$(6s)$ Rydberg molecules. $E_{\nu
0}$, $\la{R}\ra$ and $\la{\Delta R}\ra$ are the vibrational energy,
mean value and variance of the internuclear distance $R$,
respectively.}\label{tab-3}
  \begin{tabular}{c|c|c|c||c|c|c|c|c|c}
    \hline\hline
        \multicolumn{4}{c||}{$60s+6s$} &
        \multicolumn{6}{c}{$60p+6s$} \\ \hline
    $E_{\nu 0}$ (MHz) & -5.3 & -2.6 & -0.7 & $E_{\nu 0}$ (MHz) & -16.7 & -11.8 & -7.5 & -4.0 & -1.3 \\ \hline
    $\la{R}\ra$ $(a_0)$ & 5995.5 & 5930.9 & 6076.7 & $\la{R}\ra$ $(a_0)$ & 6092.2 & 6116.9 & 6149.9 & 6167.6 & 6076.6 \\ \hline
    $\la{\Delta R}\ra$ $(a_0)$ & 100.3 & 352.6 & 457.4 & $\la{\Delta R}\ra$ $(a_0)$ & 74.0 & 133.7 & 185.6 & 272.2 & 552.8 \\
    \hline\hline
  \end{tabular}
\end{table}
%

\clearpage
\clearpage
\clearpage
\clearpage

\end{document}